# Response of 3D Free Axisymmetrical Rigid Objects under Seismic Excitations

Li Yanheng    Shi Baoping

Graduate University of Chinese Academy of Sciences, Beijing 100049, China

**Abstract**  Previous studies of precariously balanced objects in seismically active regions provide important information for aseismatic engineering and theoretical seismology. They are almost founded on an oversimplified assumption: any 3-dimensional (3D) actual object with special symmetry could be regarded as a 2D finite object in light of the corresponding symmetry. To gain an actual evolution of precariously balanced objects subjected to various levels of ground accelerations, a 3D investigation should be performed. In virtue of some reasonable works from a number of mechanicians, we derive three resultant second-order ordinary differential equations determine the evolution of 3D responses. The new dynamic analysis is following the 3D rotation of a rigid body around a fixed point. A computer program for numerical solution of these equations is also developed to simulate the rocking and rolling response of axisymmetric objects to various levels of ground accelerations. It is shown that the 2D and 3D estimates on the minimum overturning acceleration of a cylinder under the same sets of half- and full-sine-wave pulses are almost consistent except at several frequency bonds. However, we find that the 2D and 3D responses using the actual seismic excitation have distinct differences, especially to north-south (NS) and up-down (UD) components. In this work the chosen seismic wave is the El Centro recording of the 18 May 1940 Imperial Valley Earthquake. The 3D outcome does not seem to support the 2D previous result that the vertical component of the ground acceleration is less important than the horizontal ones. We conclude that the 2D dynamic modeling is not always reliable.

## Introduction

In the last decades, the study on the survival of the precarious rocks perched next to a fault that produces large earthquakes has been in the ascendant (Brune, 1992; Weichert, 1994; Shi *et al.*,1996; Anooshepoor *et.al.*, 2004). Its outcome has been applied to provide constrains on peak strong ground-motion level and intensity at their locations, and thus possibly on the epicenter parameter of the earthquake and the attenuation relationship for the seismic hazard map (Brune, 1992; 1994; 1996; Brune and Whitney, 1992; Anooshepoor *et.al.*, 1999; Bell *et.al.*, 1998; Matthew *et.al.*, 2008). At the same time, the investigation on the transient response of a free-standing equipment subjected to near-source ground motions has been employed in depth to earthquake resistant designs (Housner, 1963; Yim *et.al.*, 1980; Makris and Zhang, 1999). What is the quasi-static toppling acceleration of these rocks, or what makes these motions particularly falling destructive to building objects is a focus in these researches. Almost all its findings indicate that the potential overturning accelerations is determined by not only their accidentally high earthquake forces but also the relatively long-duration acceleration pulses. Most of them are theoretically based on an ideal assumption: if the discussed objects have some symmetric, and even approximately symmetric shapes, 3D objects can be regarded as 2D planes which is only a main section of these objects. Then the problem can be mathematically reduced to 1D rotations of the planar object under various waveforms. However, 3D theoretical analysis does not manifest itself tardily due to mathematical complexity. In this study, we try to break the deadlock firstly and fortunately obtain some interesting results in the simplest 3D case----a perfectly axisymmetric cylinder under seismic excitations (Konstantinidis, D. and Makris, N, 2007). The dynamic analysis is based on the Eulerian rotation of a rigid body around a fixed point. It reveals that the cylinder with an external plus never directly drops down onto the diametrically opposite side but veers to the other side over exactly 180 degrees apart.

It seems that both of the results are equivalent in this aspect. Note that the actual veering is not always even exactly 180 degrees; nevertheless now we just concern the comparability of 2D and 3D results which the equivalence happens to present in a quite credible scale. For more complicated objects it can be still very difficult to conceive of a dynamic method of analysis, and even for a simple object the accuracy of the dynamic method has not been investigated.

Housner (1963) firstly investigated the rocking response of 2D rigid blocks due to various types of horizontal ground-motion half-sine-wave pulses. Shi *et al.*(1996) detailedly discuss the 2D toppling problem from theoretical analysis to computer simulation, and then to experimental verification with the assumptions that the block resting on a pedestal is free to rotate about either of the two supporting points *O* and *O'* in Fig.1 that no sliding occurs during the rocking motion between the block and the base. They have developed a FORTRAN program ROCKING V.1.0 (von Seggarn 2001) using a Runge-Kutta algorithm for given input values for the horizontal ($a_x$), and vertical ground acceleration ($a_z$) in that no analytical solution exists except for simple block geometries and small angles. They describe the rocking motion of 2D ideal rectangular plane using an 1D rotation of a single angle $\theta$ around a fixed point

$$I\ddot{\theta} = -mgR\sin(\alpha - \theta) - mRa_x\cos(\alpha - \theta) \qquad (1)$$

where *I* is the moment of inertia of the block about the center of mass and approximate to $4mR^2/3$ in terms of rectangular configurations with the height 2*h* and the width 2*r* (*R* measures the distance from the center of mass to *O*; $\alpha = \tan^{-1}(h/r)$). Their rough conclusions from the above-mentioned highly reduced modeling are widely accepted and referred to engineering science and seismology. (Housner, 1963; Yim *et.al.*, 1980; Brune, 1992; Weichert, 1994)

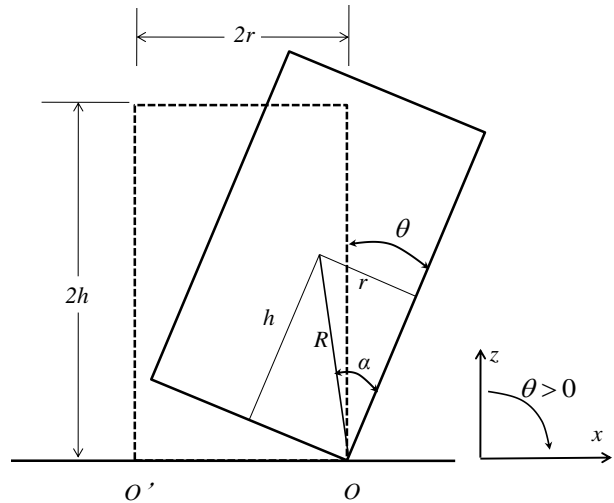

Figure 1　Schema of a 2D free-standing nonobjective rectangular plane in rocking motion

However, most of the actual objects worth of considering in engineering and seismology are three-dimensional existences in the world. When a 3D rotation is simplified into an 1D one, a rotation around a fixed point reduces one around a fixed axis and the components of mass moment of inertial in two other axes are artificially neglected. Before the feasibility and validity of the reduction is challenged, the meaning of the mass moment of inertial should be regarded. The mass moment of inertial of a rigid body measures the body's ability to resist changes in rotational speed about a specific axis. The larger the mass moment of inertial is, the smaller the angular acceleration about that axis for a given torque is. It depends on a reference coordinate and possibly contains some cross-product terms during three-dimensional motion where rotation can occur about multiple axes. The essential of this simplification is leaving out the geometric symmetry along the direction perpendicular to the rectangular plane in Fig.1. One of overt defects of this modeling serves to make the rocking processes of a cuboid and a cylinder indistinguishable. The distinctly dubious conclusion should be deliberated again by the method nearer to the facts.

In addition, during an earthquake, some freestanding equipments that are even symmetric in mass and stiffness will undergo torsional shake as well as the normal horizontal and vertical oscillations. For instance, One of the most common forms of damage in liquid tanks involves outward buckling of the bottom shell courses, a phenomenon known as "Elephant Foot Buckling". As we all know intuitively, the torque effect served as one of its destructive sources can be traced back to the nonsymmetrical features of the equipments (Housner and Outinen, 1958). In this light, what is a considerable interpretation about the torsional shake in a symmetric object? Maybe we need a simple experiment everyone can operate with your cup at hand to explain this question. Your cup on a table, when slightly tipped and released, falls to an upright position and then rolls up to a somewhat opposite tilt along the flat circular base of the cup, rather than straightly falls over onto the diametrically opposite side in 2D case. Superficially this rocking motion involves a collision when the up slaps the table before rocking up to the opposite tilt. A keen eye notices that the after-slap rising tilt is not generally just opposite the initial tilt but is veered to one side or the other. As a result, the real 3D objects exhibit distinct dynamics behavior with contrast to the oversimplified 2D ones. It is necessary to make a substantial 3D modeling using the theory of rotation around a fixed point in an ordinary 4-dimension spacetime. In this study, we have a taste of solving this problem primarily.

Otherwise, the rigid body dynamics of these objects with special symmetries have been studied at length by a number of mechanists. Reasonable reviews of such works can be found, for instance, in O'Reilly (1996) and Borisov & Mamaev (2002). Their works include not only Hamiltonian analyses of nonholonomic mechanics based on fiber bundle theory, but also complete analytical solutions to the relevant equations of motion, typically involving non-elementary functions such as the hyper-geometric. (Bloch, 1996; Zenkov, 1997). We will not use these somewhat cumbersome general solutions but will analyze only the special near-collisional motion of interest to us.

**Remodeling**

Consider a cylinder subjected to an exterior field of force—a strong ground-motion acceleration—perform a rocking and possibly rolling motion on a (*x-y*) plane without sliding. Let the cylinder with mass *m*, bottom radius *r*, and the center of mass at a height *h* from the bottom. In this case the equations of motion have the most convenient form in the body-fixed (*ξ-η-ζ*) frame of references which axes are directed parallel to the principal axes of inertia of the body and the origin is situated at the contact point *P* as Fig.2. The moment of inertia is *C* about its symmetry axis and *A* is about any axis passing through the center of mass and perpendicular to the symmetry axis.

$$A = \frac{1}{3}mr^2 + \frac{1}{4}mh^2, \quad C = \frac{1}{2}mr^2. \tag{2}$$

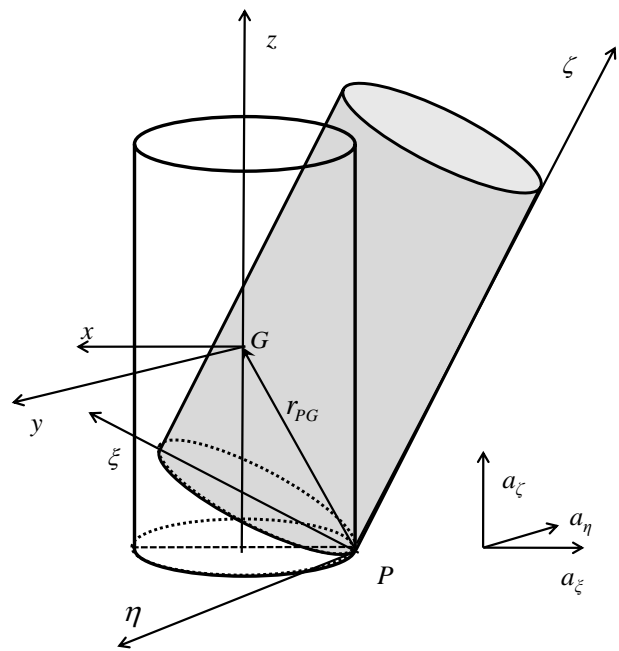

Figure 2  Definition of coordinate axes used to define the orientation of the cylinder in the

motion

The ZYZ-Euler angles are defined as the three succesice rotations. The sequence will be started by rotating the initial system of axes, *x-y-z*, by an angle $\varphi$ counterclockwise about the z axis, and the resultant coordinate system will be labeled the *x'-y'-z* axes. In the second stage the intermediate axes, *x'-y'-z*, are rotated about the y' axis counterclockwise by an angle $\theta$ to produce another intermediate set, the *x''-y'-ζ* axes. The y' axis is at the intersection of the *x-y* and *x''-y'* planes and is known as the line of nodes. Finally the *x''y'ζ* axes are rotated counterclockwise by an angle $\psi$ about $\zeta$ the axis to produce the desired *ξ-η-ζ* system of axes.

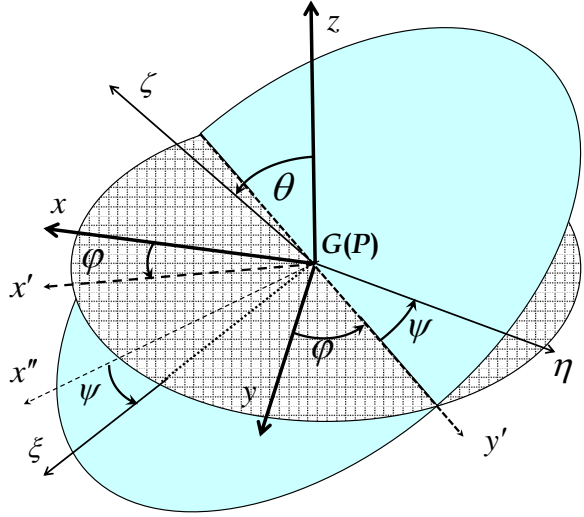

Figure 3  Definition of Euler Angles

Let ZYZ-Euler angles $\theta$, $\varphi$ and $\psi$ then denote the angle between the plane of the cylinder's bottom and the vertical axis *z*, the "heading angle" of the cylinder and "self-rotation" angle of the cylinder, respectively (Fig 3). The transform matrix between the two frames *x-y-z* and *ξ-η-ζ* is represented by

$$\begin{pmatrix} \hat{i}_x \\ \hat{j}_y \\ \hat{k}_z \end{pmatrix} = \begin{pmatrix} \cos\phi & -\sin\phi & 0 \\ \sin\phi & \cos\phi & 0 \\ 0 & 0 & 1 \end{pmatrix} \begin{pmatrix} \cos\theta & 0 & -\sin\theta \\ 0 & 1 & 0 \\ \sin\theta & 0 & \cos\theta \end{pmatrix} \begin{pmatrix} \cos\psi & -\sin\psi & 0 \\ \sin\psi & \cos\psi & 0 \\ 0 & 0 & 1 \end{pmatrix} \begin{pmatrix} \hat{i}_\xi \\ \hat{j}_\eta \\ \hat{k}_\zeta \end{pmatrix}$$

$$= \begin{pmatrix} \cos\psi\cos\theta\cos\varphi - \sin\psi\sin\varphi & -\sin\psi\cos\theta\cos\varphi - \cos\psi\sin\varphi & -\sin\theta\cos\varphi \\ \cos\psi\cos\theta\sin\varphi + \sin\psi\cos\varphi & -\sin\psi\cos\theta\sin\varphi + \cos\psi\cos\varphi & -\sin\theta\sin\varphi \\ \cos\psi\sin\theta & -\sin\psi\sin\theta & \cos\theta \end{pmatrix} \begin{pmatrix} \hat{i}_\xi \\ \hat{j}_\eta \\ \hat{k}_\zeta \end{pmatrix}$$

(3)

where $\hat{i}_x, \hat{j}_y, \hat{k}_z$ and $\hat{i}_\xi, \hat{j}_\eta, \hat{k}_\zeta$ are the two sets of unit vectors in the frames, respectively.

Thus, the unconstrained configuration space for the cylinder is $Q = R^3 \times SO(3)$. The velocities associated with the coordinates $x_G$, $y_G$, $z_G$, $\theta$, $\phi$ and $\psi$ are denoted by, $\dot{x}_G$,

$\dot{y}_G$, $\dot{z}_G$, $\dot{\theta}$, $\dot{\varphi}$ and $\dot{\psi}$, which provide the remaining coordinates for the velocity phase space *TQ*. Then the spatial angular velocity of the cylinder has the representations

$$\begin{aligned}\bar{\omega} &= \dot{\varphi}\hat{k}_z + \dot{\theta}\hat{n}_{y'} + \dot{\psi}\hat{k}_\zeta \\ &= \dot{\varphi}[(\cos\psi\hat{i}_\xi - \sin\psi\hat{j}_\eta)\sin\theta + \cos\theta\hat{k}_\zeta] + \dot{\theta}(-\sin\psi\hat{i}_\xi - \cos\psi\hat{j}_\eta) + \dot{\psi}\hat{k}_\zeta \\ &= (\sin\theta\cos\psi\dot{\varphi} - \sin\psi\dot{\theta})\hat{i}_\xi + (-\sin\theta\sin\psi\dot{\varphi} - \cos\psi\dot{\theta})\hat{j}_\eta + (\cos\theta\dot{\varphi} + \dot{\psi})\hat{k}_\zeta\end{aligned} \quad (4)$$

$\hat{n}_{y'}$ is the unite vector of the line of nodes, the *y'* axis.

Suppose an earthquake near the cylinder offers an external acceleration $\bar{a} = -a_x\hat{i}_x - a_y\hat{j}_y + a_z\hat{k}_z$, the expression of the vector of moment of momentum with respect to the point of contact *P* can be written in the following form

$$\bar{M} = \bar{r}_{PG} \times (-mg\hat{k}_z - m\bar{a}) \quad (8)$$

The Lagrangian for the problem is taken to be the kinetic energy minus the potential energy and the work done by the external force from the seismic ground motion.

$$\begin{aligned}L &= \frac{1}{2}\omega\cdot(I_P\cdot\omega) + \frac{1}{2}m(\dot{x}_G^2 + \dot{y}_G^2 + \dot{z}_G^2) - (-mg\hat{k}_z - m\bar{a})\cdot\bar{r}_{PG} \\ &= \frac{1}{2}[(A+mh^2)\sin^2\theta + (C+mr^2)\cos^2\theta + mr^2\sin^2\theta\sin^2\psi - 2mhr\sin\theta\cos\theta\cos\psi]\dot{\varphi}^2 \\ &+ \frac{1}{2}[A+mh^2+mr^2\cos^2\psi]\dot{\theta}^2 + \frac{1}{2}(C+mr^2)\dot{\psi}^2 + [mr^2\sin\theta\sin\psi\cos\psi + mhr\cos\theta\sin\psi]\dot{\varphi}\dot{\theta} \\ &+ [(C+mr^2)\cos\theta - mhr\sin\theta\cos\psi]\dot{\psi}\dot{\varphi} + mhr\sin\psi\dot{\theta}\dot{\psi} + \frac{1}{2}m(\dot{x}_G^2 + \dot{y}_G^2 + \dot{z}_G^2) \\ &+ (mg\hat{k}_z + m\bar{a})\cdot\bar{r}_{PG}\end{aligned} \quad (5)$$

Similar to the famous Chaplygin sphere, rattleback *et.al*, the constraints of the cylinder rolling and rocking on the plane are

$$\begin{aligned}\dot{x}_G &= -h\sin\theta\sin\psi\dot{\varphi} - h\cos\psi\dot{\theta} \\ \dot{y}_G &= (\cos\theta\dot{\varphi} + \dot{\psi})r - (\sin\theta\cos\psi\dot{\varphi} - \sin\psi\dot{\theta})h \\ \dot{z}_G &= r\sin\theta\sin\psi\dot{\varphi} + r\cos\psi\dot{\theta}\end{aligned} \quad (6)$$

Substitute Eq.(6) into Eq.(5), we shall write the reduced Lagrangian, and thus the Lagrange–D'Alembert–Poincaré equations. The three second order ODEs that determine the evolution of the orientation $\theta$, $\varphi$, and $\psi$ are attained by

$$\begin{pmatrix} Q_{11} & Q_{12} & Q_{13} \\ Q_{21} & Q_{22} & Q_{23} \\ Q_{31} & Q_{32} & Q_{33} \end{pmatrix} \begin{pmatrix} \ddot{\varphi} \\ \ddot{\theta} \\ \ddot{\psi} \end{pmatrix} = \begin{pmatrix} S_1 \\ S_2 \\ S_3 \end{pmatrix} \quad (7)$$

where $Q_{11} = A\sin\theta - mhr\cos\theta + mh^2\sin\theta$, $Q_{12} = 0$, $Q_{13} = -mhr$,

$Q_{21} = Q_{23} = 0$, $Q_{22} = -mr^2 - mh^2 - A$,

$Q_{31} = C\cos\theta + mr^2\cos\theta - mrh\sin\theta$, $Q_{32} = 0$, $Q_{33} = C + mr^2$,

$S_1 = (C - 2A - 2mh^2)\dot\varphi\dot\theta\cos\theta + C\dot\theta\dot\psi - 2mhr\dot\varphi\dot\theta\sin\theta - ha_y$

$S_2 = (C - A + mr^2 - mh^2)\dot\varphi^2\sin 2\theta/2 + mhr\dot\psi^2$ (8)

$\quad + mhr\cos 2\theta\,\dot\varphi^2 + (C + 2mr^2)\sin\theta\dot\varphi\dot\psi$

$\quad + m(g + a_z)(r\cos\theta - h\sin\theta) + 2mhr\cos\theta\dot\varphi\dot\psi$

$\quad + ma_x(r\sin\theta + h\cos\theta)$,

$S_3 = C\sin\theta\dot\varphi\dot\theta + 2mr(r\sin\theta + h\cos\theta)\dot\varphi\dot\theta + a_y r$.

In order to simplify the intricate and fallible calculation, we will only consider a trivial case of the ground acceleration with a single *x*-component or *x*- and *y*-components, namely, the *x*–coordinate being oriented by the external acceleration's exposure. Meanwhile, Cushman and Duistermaat (2006) recently noticed such veering when a flat disk with rolling boundary conditions is dropped nearly flat. On account of the obvious symmetries in the physics, the angles $\varphi$ and $\psi$ do not affect the dynamics directly, but only through their rates $\dot\varphi$ and $\dot\psi$. Srinivasan and Ruina (2007) extend these rolling disk results to arbitrary axisymmetric bodies. Thus in our calculation we could safely assign to $\varphi$ and $\psi$ infinitesimals and switch insensibly before a so-called near-collisional falling-down. According to Euler-Jacobi Theorem,

$$\frac{d\vec{J}}{dt} = \vec{M} \quad (9)$$

we also should deduce the three second order ODEs, Eq(7-8).

## Results

### Typical Example

At first, we only choose a set of half-sine pulses of acceleration as the input motion. The equation of motion (Eq.1) of a rocking body subjected to a half-sine-wave ground acceleration of $a_x(t)$ is

$$a_x(t) = \begin{cases} -A_x \sin(\omega t + \phi) & -\frac{\phi}{\omega} \leq t \leq \frac{\pi - \phi}{\omega} \\ 0 & otherwise \end{cases} \quad (10)$$

Here $A_x$ is the amplitude of the input waveform with frequency $\omega$. We also employ the initial condition which requires the base acceleration is larger enough to initiate rocking of the block at the instant *t*=0 (*e.g.* Brune, 1992), the initial phase $\phi$ should satisfy

$$\phi = \sin^{-1}\left(\frac{\alpha g}{A_x}\right), \quad g=9.81\text{m/s}^2 \tag{11}$$

We set $\dot{\theta}(0) = \varepsilon$, $\dot{\varphi}(0) = \varepsilon$, $\dot{\psi}(0) = \varepsilon$, $\theta(0) = \varepsilon$ and $\varphi(0) = \psi(0) = 0$, where $\varepsilon$ is a small quantity. One of the numerical solutions of Eqs.(7-8) is displayed in Figure 4, where the plots of three Euler angles and their first derivations versus time present the motion of the cylinder under a sin-wave plus in the time interval [0, 10s]. Physical parameters $m=1$, $\alpha = 15^0$, $h = 0.5$ in consistent SI units in the calculation, and the frequency $\omega = 0.5\,\text{Rad/s}$. $A_x = 0.313g$, the behavior of angle $\theta$ seems analogous to the one in 2D case, rocking up from one point to the opposite one. The other two remain nearly constant and their velocities keep zero in each segment of motion until the bottom face falls down to the ground $\theta=0$. As Srinivasan and Ruina (2007) discussed, the velocities $\dot{\varphi}$ and $\dot{\psi}$ are very close to being equal and opposite. Clearly point P runs back and forth as the evolution of the Euler angles $\theta$ and $\varphi$. The position of the contact point P ($x_P, y_P, z_P$) on the ground relative to the center of mass G ($x_G, y_G, z_G$) is given by the following equations:

$$\begin{aligned} x_P &= x_G - r\cos\theta\cos\varphi + h\sin\theta \\ y_P &= y_G - r\sin\varphi \\ z_P &= z_G - r\sin\theta\cos\varphi - h\cos\theta \end{aligned} \tag{12}$$

At the twinkling time when the cylinder is almost vertical $\theta=0$,

$$\begin{aligned} x_P &= x_G - r\cos\varphi \\ y_P &= y_G - r\sin\varphi \\ z_P &= z_G - h = 0 \end{aligned} \tag{13}$$

and a rapid finite change in $\varphi$ by a matched asymptotic calculation,

$$\Delta\varphi_{turn} \approx \pi\sqrt{\frac{A+mh^2}{A+mr^2+mh^2}} \tag{14}$$

because the case in our study belongs to small-$\theta$ dynamic regime (Srinivasan and Ruina, 2007).

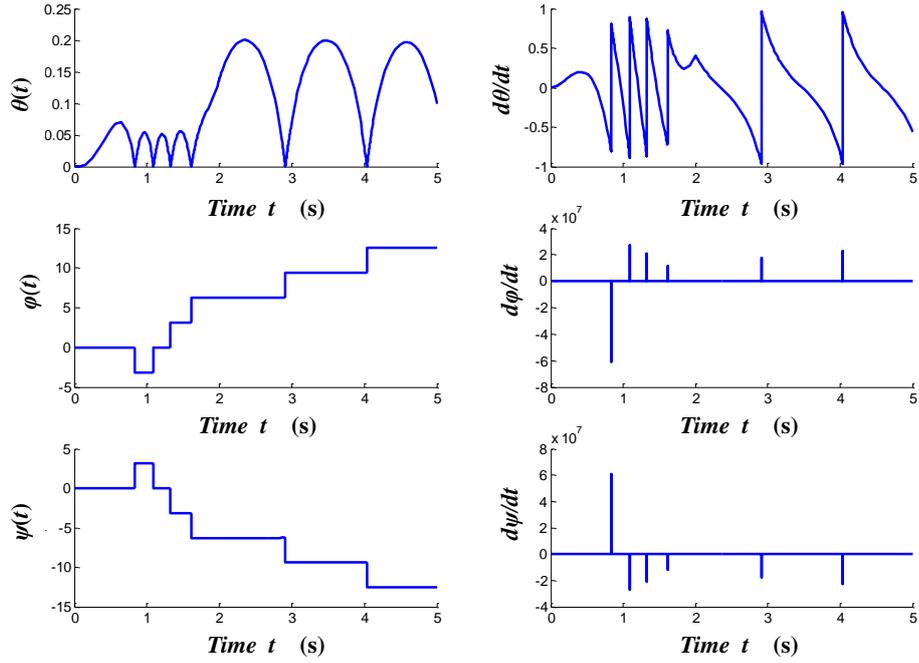

Figure 4  Results of a typical simulation of a cylinder's face almost but not quite falling flat on the ground under a half-sine pulse with amplitude $A_x=0.313g$ and frequency $\omega=0.5$ Rad/s. Physical parameters $g = 9.81$m/s$^2$; $r =0.5$m; $\alpha =15^o$ in consistent SI units. Initial conditions $\dot\varphi(0)= 5\times 10^{-5}$, $\dot\psi(0)= 5\times 10^{-5}$, $\dot\theta(0)=5\times 10^{-5}$, $\theta(0)=5\times 10^{-5}$, $\varphi(0)=0$, $\psi(0)=0$. All the angles are in radians.

Substitution Eq.(14) with $\alpha=15^0$, $h=0.5$, we can get $\Delta\varphi_{turn} \approx \pi$, which then reveals from Eq.(13) that the contact point moves to point $P'$, the antipodal point of the old fixed point $P$ with respect to the base-circle centre. Subsequently the cylinder rotates around the new fixed point $P'$, and thus it rocks down and up periodically between point $P$ and $P'$ seemingly like 2D pendulum-like process. The evolution of angle $\theta$ spontaneously has a little attenuation, where the ratio of the maximum values of $\theta$ between all the two nearest periods is as much as 0.95 around. It means that in this process the total energy of the cylinder slightly decreases duratively. One of possible mechanisms is that when the cylinder's face almost but not quite falling flat on the ground, the friction force between the bottom rim and the ground serves to the rates of angle $\dot\varphi$ and $\dot\psi$ undergoing an abrupt jump from a huge value to zero. Perhaps it is responsible for the losing energy. As it turns out, in the computer code we need not refer ourselves to a nonphysical parameter----attenuation ratio controlling the evolution of angle $\theta$, but 2D simulation need.

**Comparison I**

For comparison, we demonstrate the different rocking motions of the freestanding cylinder based on 2D and 3D dynamic simulations under the same condition----a full-sine pulse with 1Hz and its amplitude $A_x=0.35g$ is presented by Fig. 5 and Fig. 6. Obviously, 3D object wobbles with a small angle continuously while 2D one topples down after a one-period pendulum-like rock. In the one-period interval of 1.0 Rad/s sine wave, 3D object finishes three-period pendulum-like rock with maximum precession angle $\theta \approx 2.5°$ which is the same as the one in the first 2D period.

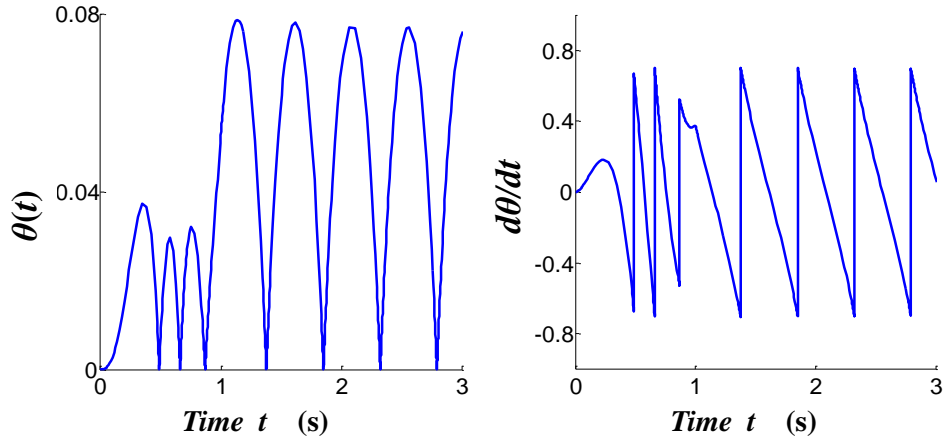

Figure 5  Plots of domain Euler angle $\theta$ and its rate of the cylinder subjected to a full-sine-wave ground acceleration with amplitude $A_x=0.35g$ and frequency $\omega=1.0$ Rad/s as a function of time t. Initial conditions are the same as Fig. 4.

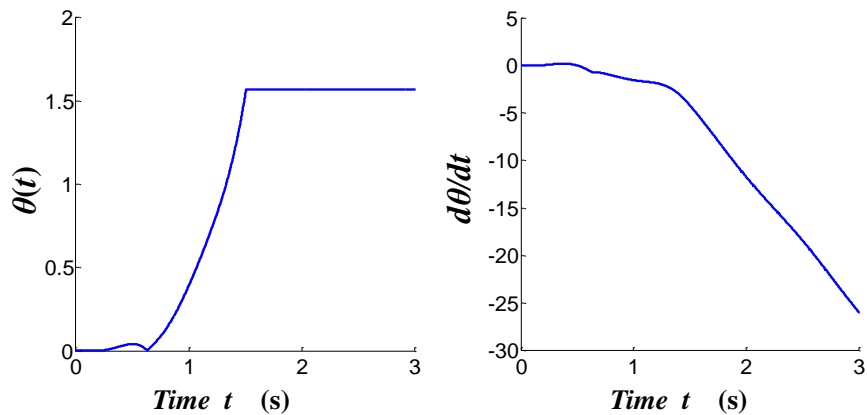

Figure 6  Results of a typical simulation of rocking of a 2D rectangular plane replacing a 3D cylinder subjected to a full-sine-wave ground acceleration with amplitude $A_x=0.35g$ and frequency $\omega=1.0$ Rad/s.

**Comparison II**

Fig. 7 shows that a comparison on the minimum overturning accelerations of 2D objects and 3D ones under the same external conditions. The two sets of curves with respect to different frequencies of half- and full-sine-waves as an ingredient of real earthquake shock, on the whole, are almost coincident except at several full-sine-wave frequency bonds, such as 0.5-3.0 Rad/s, >5.5 Rad/s. In terms of half-sine waves, the two curves overlap in all regions in Fig.7. Especially, it is interesting that at about 0.5-3.0 Rad/s 3D objects fall down quite more lately than 2D objects under a serial of sine-plus with uniformly increasing accelerations. That is to say, 3D objects are more "lodging resistant". Whereas the situation dramatically becomes converse when the frequency is more than 5.5Hz, the largest variation between these critical accelerations required to topple this cylinder is considerably equal to 0.5g.

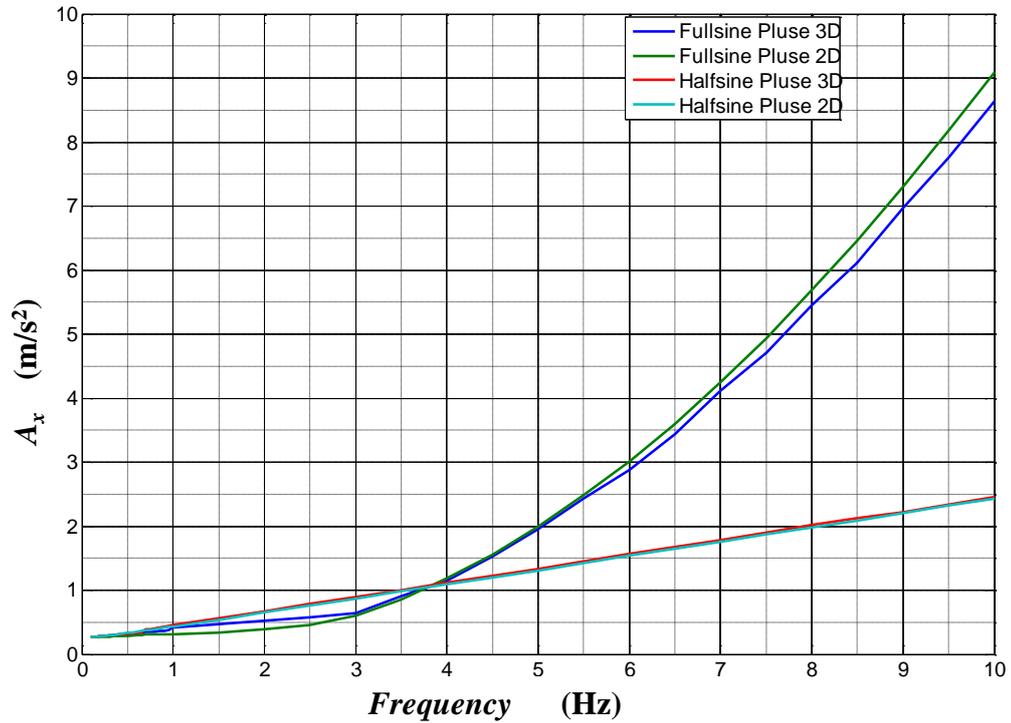

Figure 7　The minimum toppling amplitude $A_x$ of rigid cylinders subjected to full (dashed line) and half-sine-wave acceleration pulses (solid line) are plotted as functions of frequency $\omega$. Here, $A_x$ and $\omega$ are the amplitude and frequency of the sinusoidal input motion, respectively; $g$ is the acceleration due to gravity.

**Comparison III**

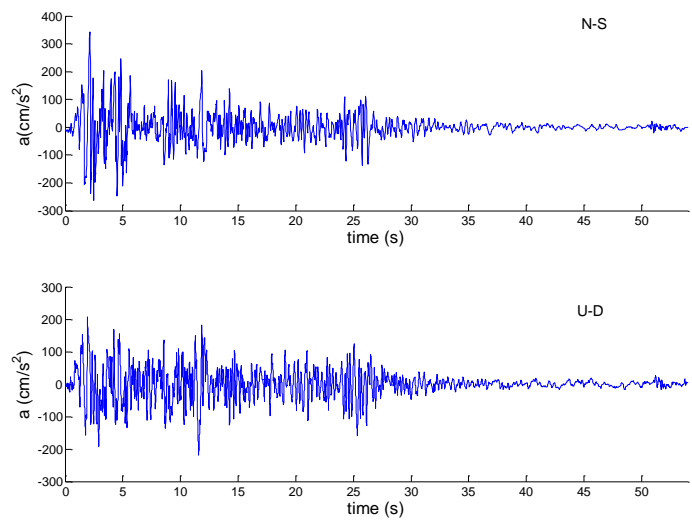

Figure 8　The north-south and vertical components of the E1 Centro recording of the 1940 Imperial Valley earthquake.

We validated our numerical code by comparing the 2D and 3D responses of axisymmetrical blocks with different aspect ratios ($h/r$) subject to the El Centro ground-motion recording of the 18 May 1940 Imperial Valley earthquake (Fig 8). Fig. 9 and Fig. 10 indicate the two regimes of rocking and overturning of different scale objects. In order to find out their boundary at each height $h$, we judge whether the objects rock or fall down by calculating the peak value of Euler angle $\theta$ at 10 equidistant points spreading in certain scope of the width $r$ (in Fig.9, $2h/25 \leqslant r \leqslant 7h/20$, and in Figure 10, $h/10 \leqslant r \leqslant h/5$). Fig.9 displays the responses to only the north-south component using the 2D and 3D modeling respectively, which implies several differences between them at the height of 2.5-3.0m and 5.5-6.0m. The 2D and 3D responses to north-south and vertical components are indicated in Fig.10 where shows some significant differences between them. It suggests that the two other components of inertial matrix neglected in the 2D modeling possibly play an important role to determine the responses of axisymmetric objects under real earthquake ground motion. With respect to the height of 1-3m, the 2D modeling predicts larger overturning width than 3D does, and for the height of 4m-6m, the result is converse. The 3D result shows that the division line between rocking and overturning regions is approximately a straight line with a slope of 1/7. Comparing Fig. 9 and Fig. 10, the two figures without and with consideration of the vertical component indicate obvious distinction. Thus we should reconsider the previous 2D result that the vertical component of the ground acceleration is less important than the horizontal ones. These figures also illustrate a strong dependence between the stability and the aspect ratio $(h/r)$ and size (R). In general, the cylinders of smaller aspect ratio and larger size are more stable.

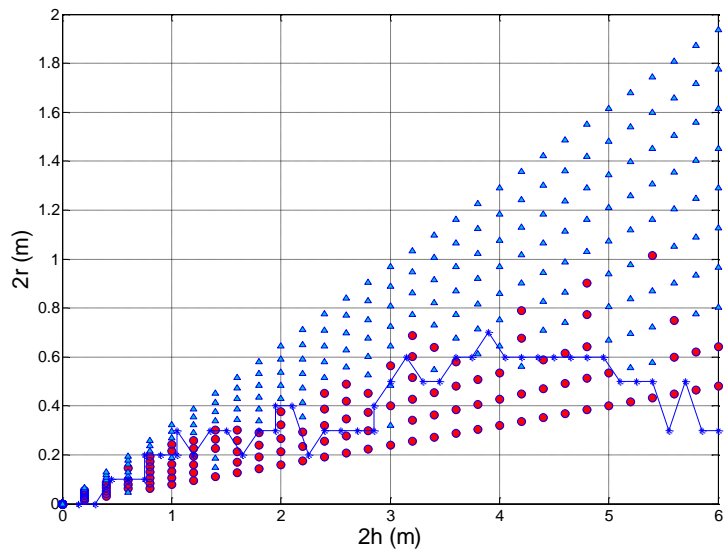

Figure 9  Comparison of 2D and 3D dynamic stable analysis of rectangular blocks with various aspect ratios $(h/r)$ subjected to the NS component of the E1 Centro strong-motion recording of the 1940 Imperial Valley earthquake. The broken line connected by ∗ is the boundary of rocking and overturning regions using 2D modeling. 3D Rocking is indicated by triangle, and overturning by circle.

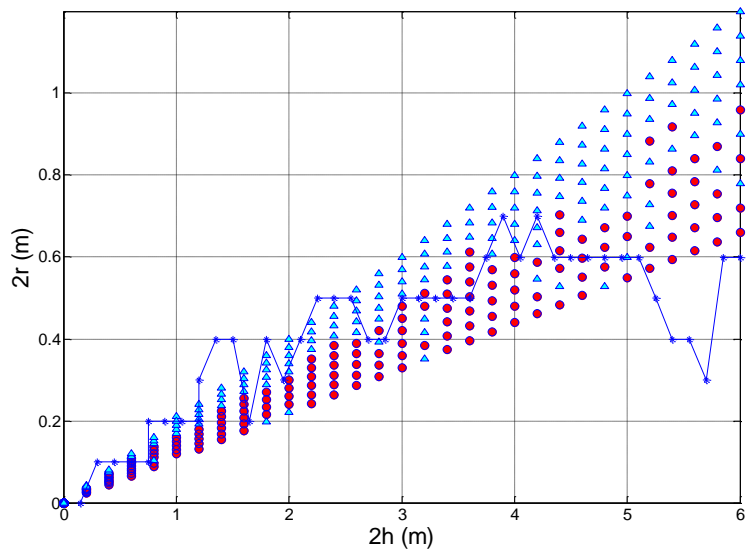

Figure 10   Comparison of 2D and 3D dynamic stable analysis of rectangular blocks with various aspect ratios *(h/r)* subjected to Both of the NS and UD components of the E1 Centro strong-motion recording of the 1940 Imperial Valley earthquake. The symbols are the same as Figure 9.

**Concluding Remarks**

This work tries to model the response of a 3D cylinder subjected to a set of sine waves as some components of an actual seismic wave, and most of its results in some terms agree with the former works in the 2D case. However, when referred to actual seismic excitation---- the El Centro recording of the 18 May 1940 Imperial Valley Earthquake, the 2D and 3D results have distinct differences, especially to NS and UD components as an input. The 3D outcome seems to question the 2D result that the vertical component of the ground acceleration is less important than the horizontal ones. Our conclusion suggests that 2D dynamic modeling is not always reliable and enough precise to offer some reasonable knowledge to seismic research in regard to ground-motion response of a free-standing object with special symmetries. In words, 3D dynamic analysis of the seismic response is necessary and more valid. The further improvement we expect based on this study includes modeling other general objects with various shapes. Maybe the evaluation of three Euler angles with less simplification should be taken into account and remains to be explored in detail. Besides, the mechanic analysis is also needed, due to the fact that the torque when rolling is concernful and often discussed in the aseismic design of unanchored liquid-tanks.